\documentstyle[epsfig,floats,prl,aps]{revtex}
\input psfig.sty
\begin{document}
\twocolumn
%\draft
\title{The Dynamics of Component Separation in a Binary Mixture of
  Bose-Einstein Condensates}
\author{D.~S. Hall, M.~R. Matthews, J.~R. Ensher, C.~E.
  Wieman, and E.~A. Cornell\cite{qpdNIST}}
\address{JILA, National Institute of Standards and Technology and
  Department of Physics, University of Colorado,\\ Boulder, Colorado
  80309-0440}
\date{April 2, 1998}
\maketitle
\begin{abstract}  
  We present studies of the time-evolution of a two-component system
  of Bose-Einstein condensates (BEC) in the $\left|F=1,m_f=-1\right>$
  and $\left|2,1\right>$ spin states of ${}^{87}$Rb. The two
  condensates are created with a well-defined relative phase and
  complete spatial overlap. In subsequent evolution they undergo
  complex relative motions that tend to preserve the total density
  profile. The motions quickly damp out, leaving the condensates
  in a steady state with a nonnegligible (and adjustable) overlap region.
\end{abstract}

Since its realization in dilute atomic
gases~\cite{Anderson95,Davis95,Bradley97a}, Bose-Einstein condensation
(BEC) has afforded an intriguing glimpse into the macroscopic quantum
world. Attention has recently broadened to include exploration of
systems of two or more condensates, as realized in a magnetic trap in
rubidium~\cite{Myatt97} and subsequently in an optical trap in
sodium~\cite{Stamper-Kurn98}. Theoretical treatment of such systems
began in the context of superfluid helium
mixtures~\cite{Khalatnikov57q} and spin-polarized
hydrogen~\cite{Siggia80}, and has now been extended to BEC in the
alkalis~\cite{Ho96q,Pu98a,Goldstein97q}.

The first experiments involving the interactions between
multiple-species BEC were performed with atoms evaporatively cooled in
the $\left|F=2,m_f=2\right>$ and $\left|1,-1\right>$ spin states of
${}^{87}$Rb~\cite{Myatt97}. These experiments demonstrated the
possibility of producing long-lived multiple condensate systems, and
that the condensate wavefunction is dramatically affected by the
presence of interspecies interactions. In this Letter, we report
results from initial studies of simultaneously trapped BECs in the
$\left|2,1\right>$ and $\left|1,-1\right>$ states of ${}^{87}$Rb
(denoted hereafter as $\left|2\right>$ and $\left|1\right>$,
respectively). The two states are completely distinguishable since the
hyperfine splitting is much larger than any other relevant energy
scale in the system. We produce arbitrary superpositions of
$\left|1\right>$ and $\left|2\right>$ that begin with a well-defined
relative phase, spatial extent, and ``sag'' --- the position at which
the magnetic trapping forces balance gravity for each state. The fine
experimental control of this double condensate system permits us to
study its subsequent time-evolution under a variety of interesting
conditions, most notably those in which there remains substantial
overlap between the two states.

The apparatus and general procedure we use to attain BEC in Rb are
identical to those of our previous work~\cite{Matthews98} and will be
reviewed here but briefly. We use a double magneto-optical trap system
to load roughly $10^9$ $\left|1\right>$ atoms into a time-averaged,
orbiting potential (TOP) magnetic trap~\cite{Petrich95}. The atoms are
magnetically compressed and evaporatively cooled~\cite{Anderson95} for
30~s until they form a condensate of approximately $5 \times 10^5$
atoms with no noticeable non-condensate fraction (we estimate that
$>75$\% of the entire gas is in the condensate). After completion of
the evaporation cycle, the magnetic trap is ramped adiabatically to
various bias fields and spring constants for the subsequent
experiments.

\begin{figure}
\begin{center}
\psfig{figure=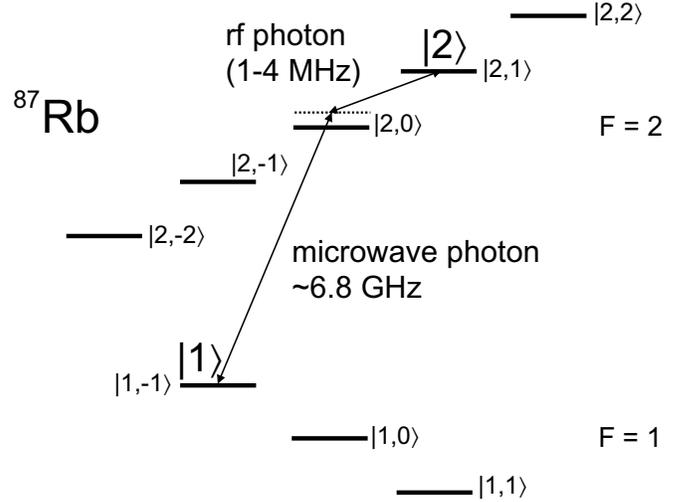,width=\linewidth,clip=}
\end{center}
\caption{A schematic of the ground-state hyperfine levels
  (\protect{$F=1,2$}) of \protect{${}^{87}$Rb}, shown with a splitting
  due to an applied magnetic field. The two-photon transition is
  driven between the \protect{$\left|1,-1\right>$}
  (\protect{$\left|1\right>$}) and \protect{$\left|2,1\right>$}
  (\protect{$\left|2\right>$}) states.}
\label{states}
\end{figure}

The double condensate system is prepared from the single
$\left|1\right>$ condensate by driving a two-photon
transition~\cite{Matthews98} consisting of a microwave photon near
6.8~GHz and a radiofrequency (rf) photon of 1--4~MHz, depending on the
Zeeman splitting (Fig.~\ref{states}). As in \cite{Matthews98}, we are
able to transfer quickly any desired fraction of the atoms to the
$\left|2\right>$ state by selecting the length and amplitude of the
two-photon pulse. The two condensates are created with identical
density distributions, after which they evolve and redistribute
themselves for some time $T$.  We then turn off the magnetic trap and
allow the atoms to expand for 22~ms for imaging.

We selectively image the densities of either of the two states ($n_1$
and $n_2$) or the combined density distribution ($n_T$) by changing
the sequence of laser beams applied to the condensates for
probing~\cite{Matthews98}. Since the expansion and imaging are
destructive processes, each image is taken with a different
condensate; the excellent reproducibility of the condensates permits
us to study the time-evolution of the system by changing the time $T$.
We subsequently use a least-squares algorithm to reconstruct the
relative positions of the condensates from the images of $n_1$, $n_2$,
and $n_T$ at each time $T$ --- a task made necessary by shot-to-shot
jitter in the image positions on the CCD-array detector.

The evolution of the double condensate system, including the release
from the trap and subsequent expansion~\cite{Anderson95,Holland96}, is
governed by a pair of coupled Gross-Pitaevskii equations for
condensate amplitudes $\Phi_i$:
\begin{equation}
i\hbar\frac{\partial\Phi_i}{\partial t} = \left(-\frac{\hbar^2\nabla^2}{2m} +
V_i + U_i + U_{ij}\right)\Phi_i
\label{gpe}
\end{equation}
where $i,j=1,2~(i \ne j)$, $V_i$ is the magnetic trapping potential
for state $i$, the mean-field potentials are $U_i =
4\pi\hbar^2a_i|\Phi_i|^2/m$ and $U_{ij} =
4\pi\hbar^2a_{ij}|\Phi_j|^2/m$, $m$ is the mass of the Rb atom, and
the intraspecies and interspecies scattering lengths are $a_i$ and
$a_{ij}$. In the Thomas-Fermi limit, the condensate density
distributions are dominated by the potential energy terms of
Eq.~(\ref{gpe}). Consequently, the expanded density distributions
retain their spatial information and emerge with their gross features
(such as the relative position of the condensates) intact.

The similarity in scattering lengths $a_1$, $a_2$, and $a_{12}$ implies
that the total density $n_T$ will not change significantly from its
initial configuration even though the two components may redistribute
themselves dramatically during the evolution time $T$. In ${}^{87}$Rb,
the scattering lengths are known at the 1\% level to be in the
proportion $a_1:a_{12}:a_2::1.03:1:0.97$, with the average of the
three being 55(3)~\AA~\cite{Burke_PVT,Matthews98}.  The
near-preservation of the total density $n_T$ can be approached
theoretically by deriving from Eq.~(\ref{gpe}) the hydrodynamic
equations of motion~\cite{ZarembaXX} for $n_T$ and evaluating them in
the limit that the fractional differences between the scattering
lengths are small. The pressures that tend to redistribute $n_T$ must
also be small. A similar argument pertains if the minima of the
trapping potentials $V_1$ and $V_2$ are displaced from one another
(see below) by a distance that is small compared to the size of the
total condensate; once again, the effects on the equilibrium
distribution of the individual components may be profound but the
total density should remain largely unperturbed~\cite{Esry_PVT}.

The rotating magnetic field of the TOP trap~\cite{Petrich95} gives
rise to a subtle behavior that permits us to displace the minima of
the trapping potentials $V_1$ and $V_2$ with respect to one
another~\cite{Bohn_PVT,Hall98}. In the rotating frame, the two states
see two different magnetic fields as a function of the bias field
rotation frequency and sense of rotation (as well as the strengths of
the bias and quadrupole fields). By adjusting these parameters, we
can change the sign of the relative sag or cause it to
vanish~\cite{trapnote} while preserving (to first order) the same
radial ($\nu_r$) and axial ($\nu_z=\sqrt{8}\nu_r$) trap oscillation
frequencies.

\begin{figure}
\begin{center}
\psfig{figure=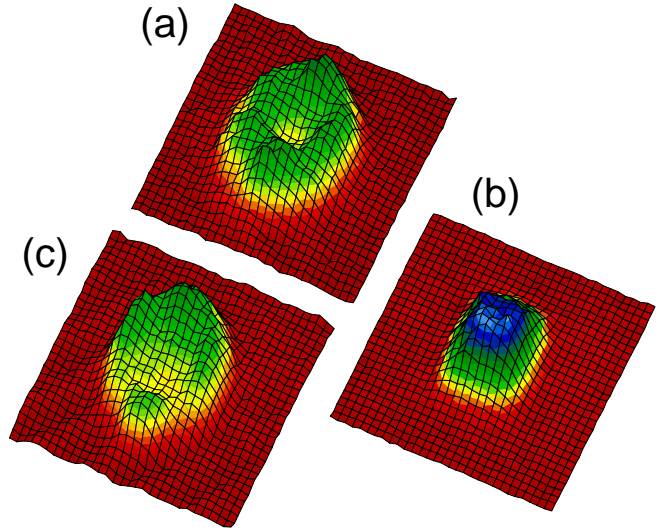,width=\linewidth,clip=}
\end{center}
\caption{(Color) (a) The
  image of the \protect{$\left|1\right>$} condensate exhibits a
  ``crater,'' corresponding to a shell in which the
  \protect{$\left|2\right>$} atoms (b) reside. For this trap,
  \protect{$\nu_z = 47$}~Hz with zero relative sag. By changing the
  strength of the magnetic quadrupole field we can introduce a nonzero
  relative sag, which shifts the location of the ``crater'' (c). (Each
  square in this post-expansion image is \protect{$136~\mu$}m on a side.)}
\label{crater}
\end{figure}

In a first experiment, we choose a trap that has zero relative sag
($\nu_z=47$~Hz) and transfer 50\% of the atoms to the $\left|2\right>$
state with a $\sim400~\mu$s pulse. When $T=30$~ms, we observe a
``crater'' in the image of the $\left|1\right>$ atoms
(Fig.~\ref{crater}a).  The ``crater'' corresponds to a region occupied
by the $\left|2\right>$ atoms (Fig.~\ref{crater}b), indicating that
the $\left|1\right>$ atoms have formed a shell about the
$\left|2\right>$ atoms. This is consistent with the theoretical
observation that it is energetically favorable for the atoms with the
larger scattering length ($\left|1\right>$) to form a lower-density
shell about the atoms with the smaller scattering length
($\left|2\right>$)~\cite{Pu98a}. At longer times the condensates
separate from one another radially~\cite{offsetnote}.

\begin{figure}
\begin{center}
\psfig{figure=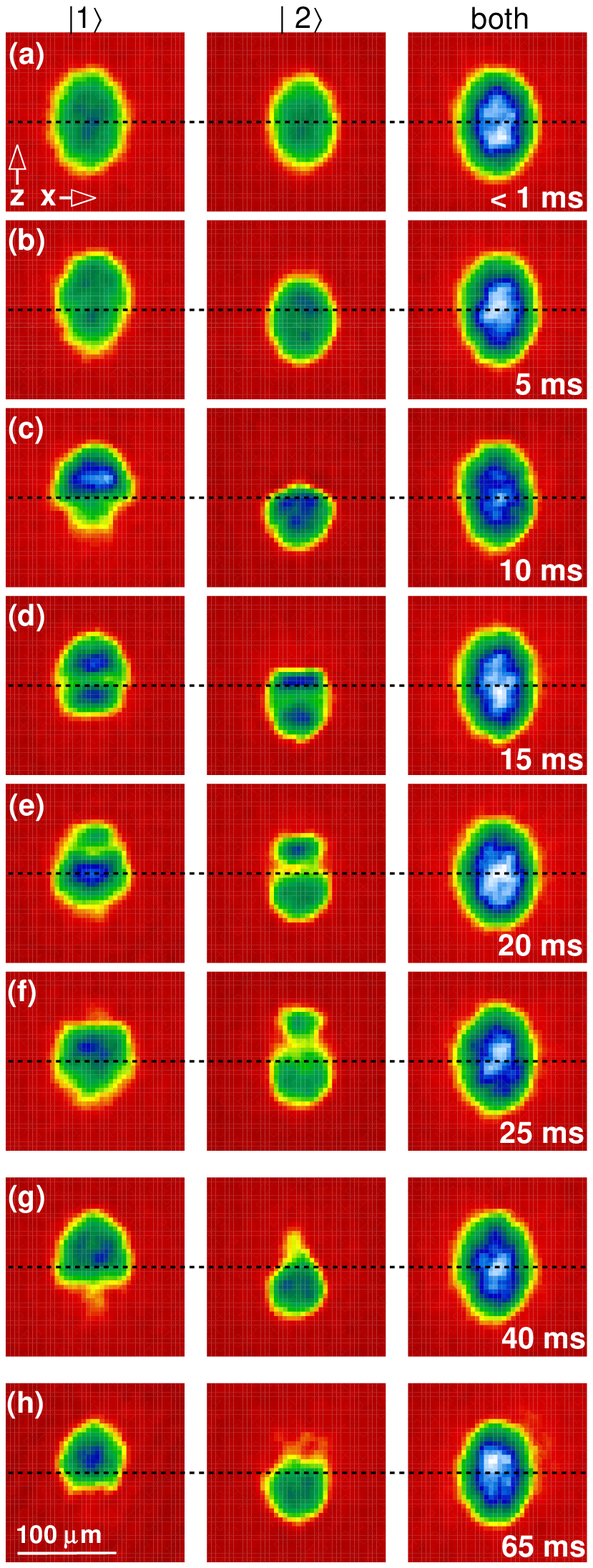,width=0.98\linewidth,clip=}
\end{center}
\end{figure}
\begin{figure}
\caption{(Color) Time-evolution of the double-condensate system with a
  relative sag of \protect{$0.4~\mu$}m (3\% of the width of the
  combined distribution prior to expansion) and a trap frequency
  \protect{$\nu_z = 59$~Hz}.}
\label{evolve}
\end{figure}
    
\begin{figure}
\begin{center}
\psfig{figure=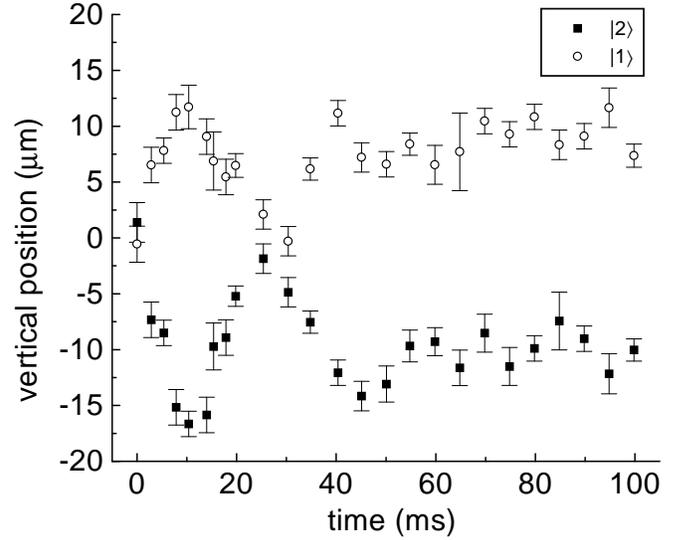,width=\linewidth,clip=}
\end{center}
\caption{The relative motion of the centers-of-mass of the two
  condensates under the same conditions as those in
  Fig.~\protect{\ref{evolve}}.}
\label{mode}
\end{figure}

In order to explore the boundary between the two condensates, we
perform a series of experiments in a trap in which we displace the
trapping potentials such that the minimum of $V_2$ is $0.4~\mu$m lower
than that of $V_1$, or approximately 3\% of the (total) extent of the
combined density distribution in the vertical direction. The
subsequent time-evolution of the system is shown in Figs.~\ref{evolve}
and~\ref{mode}. The two states almost completely separate
(Fig.~\ref{evolve}a-c) after 10~ms; they then ``bounce'' back until at
$T=25$~ms the centers-of-mass are once more almost exactly
superimposed (Fig.~\ref{mode}), although a distinctive (and
reproducible) vertical structure has formed (Fig.~\ref{evolve}d-e-f).
By $T=65$~ms, the system has apparently reached a steady state
(Fig.~\ref{evolve}g-h, Fig.~\ref{mode}) in which the separation of the
centers-of-mass is 20\% of the extent of the cloud. From these images
we observe: (i) the fractional steady-state separation of the expanded
image is large compared to the fractional amount of applied symmetry
breaking, as we expect for a repulsive interspecies potential; (ii)
the placid total density profile (rightmost column of
Fig.~\ref{evolve}) betrays little hint of the underlying violent
rearrangement of the component species; and (iii) the component
separation is highly damped, although it is not yet certain what
mechanism~\cite{Ruprecht95q} is responsible. With respect to the
damping, the excitation is in no sense small and may therefore be
poorly modeled by theories that treat the low-lying, small-amplitude
excitations~\cite{Goldstein97q} of double condensates.

\begin{figure}[t]
\begin{center}
\psfig{figure=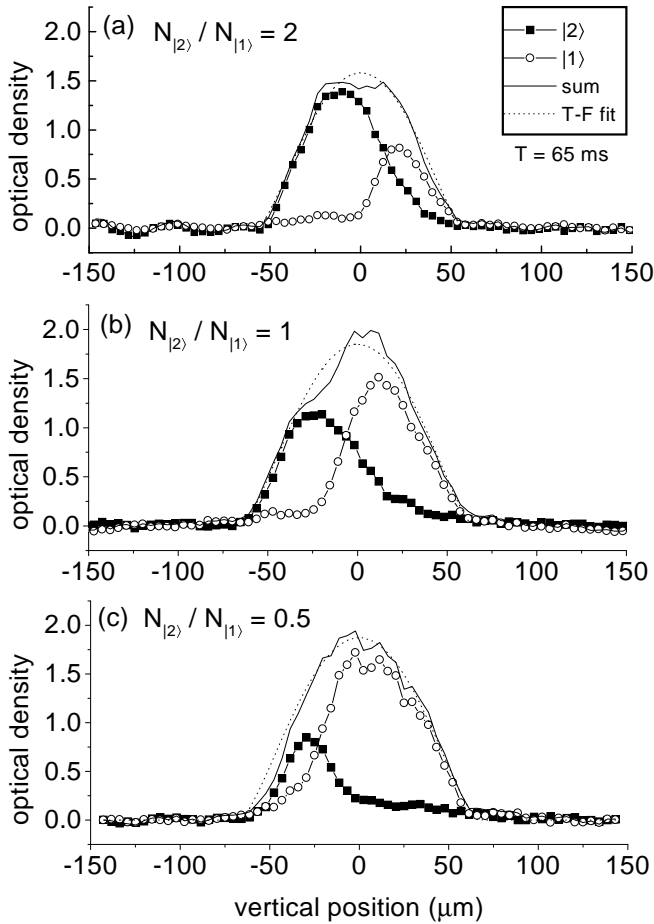,width=\linewidth,clip=}
\end{center}
\caption{Vertical cross-sections of the density profiles at
  \protect{$T=65$}~ms for different relative numbers of atoms in the two
  states. The combined density distribution (solid line) is shown for
  comparison to the Thomas-Fermi parabolic fit (dashed line). The trap
  parameters are the same as those in Fig.~\protect{\ref{evolve}}. }
\label{stripes}
\end{figure}

Finally, we show the optical density as a function of relative number
and position on the condensate vertical axis in order to better
appreciate the amount of overlap between the two states at $T=65$~ms
(Fig.~\ref{stripes}), which remains substantial despite the underlying
separation. Each plot is averaged across a $\sim 14~\mu$m wide
vertical cut through the centers of the two condensates. From the
overlap shown, one could determine the magnitude of the interspecies
scattering length $a_{12}$ by comparison to numerical solutions of the
Gross-Pitaevskii equations (\ref{gpe}) for our trapping conditions.
Such a calculation is beyond the scope of the present work.

In related work, we have read out the relative quantum phase of the
condensates in the overlap region with a second two-photon pulse,
making use of Ramsey's method of separated oscillatory fields. We
expect the presence of a weak cw two-photon drive, resonant in the
overlap region, to result in phase-locking and phase-sensitive
currents analogous to the Josephson effect in
superconductors~\cite{Josephson62q}. We will discuss the
phase-evolution of the condensates in a future publication.

We gratefully acknowledge useful conversations with the other members
of the JILA BEC collaboration, in particular those with Chris Greene
and John Bohn. This work is supported by the ONR, NSF, and NIST.

%\bibliographystyle{prsty}
%\bibliography{prl2,bec}

\begin{thebibliography}{10}
\bibitem{offsetnote}
We believe this effect to be related to
  asymmetries in the trapping fields.
\end{thebibliography}

\end{document}